\documentclass[12pt,a4paper,english]{article}
\usepackage{babel,amssymb,amsmath,graphicx,hyperref}

\begin{document}

\begin{center}
{\LARGE \textbf{Image of a black hole \\ illuminated by a parabolic screen} }\\~\\
{\large Elena V.~Mikheeva$^1$,  Serge~Repin$^{1,2}$,  Vladimir N.~Lukash$^{1}$}
\end{center}
\begin{center}
    \textit{$^1$Astro Space Center of the P.N.~Lebedev Physical Institute of the Russian Academy of Sciences, 
                         Profsoyuznaya st. 84/32, Moscow, 117997, Russia}
\end{center}
\begin{center}
    \textit{$^2$Physics and Mathematics School No. 2007,
     st. Gorchakova, 9, k.~1, Moscow, 117042, Russia}
\end{center}

\begin{abstract}
         An analytical model of a parabolic screen illuminating a black hole is proposed. This allows to avoid naturally the appearance of edge effects associated with photons moving along the plane of the screen. The temperature distribution along the radius of the screen corresponds to that for a relativistic disk (Novikov-Thorne disk).  It is shown that the structure of the emerging black hole shadow differs significantly from the case when the photon source is a remote screen, since in the model considered, the photons subjected to strong gravitational lensing of the black hole are emitted by the "back side" of the screen, which would not be visible in the absence of a black hole.  In the thin screen approximation, the shadow of a Schwarzschild black hole has been constructed in cases when the angle between the axis of symmetry of the illuminating screen and the direction towards the observer is 5, 30, 60, and 80 degrees. For the Kerr black hole, images are shown for angles of 60 and 80 degrees.
\end{abstract}

\section{Introduction}

         Modeling the shadows of black holes (BHs) is an important problem in modern astrophysics. 
         
         Initially, such modeling was mainly illustrative, making the predictions of the General Relativity clearer, but as the observational base developed, research interest shifted towards constructing the BH shadows, taking into account the instrumental restrictions: noise, resolution, observation frequencies, etc. At the same time there was a development of ideas about  the physical conditions in accretion disks and jets that illuminate BHs.

         Numerous factors make the tasks of building and restoring an image of a BH shadow quite difficult. To date, only two BHs have been reconstructed based on observational data: M87* \cite{m87} and SgrA* \cite{sgra}. This short list is planned to be supplemented by active supermassive BHs with large angular sizes, which will be observed at higher frequencies ($\simeq 350$~GHz) and at larger baselines. The last condition will be achieved by bringing the telescope out of the Earth.

       The dependence of the BH shadow image on many parameters (and the possibility of recovering it) is of interest in such shadow modeling, in which the number of these parameters is minimized in order to reveal the influence of a limited number of qualities on the image. The main factors influencing the shape of the BH shadow are the strong gravitational lensing of photons in the BH field and the shape of the photon source that illuminates it.

       Previously, we have already considered the problem of modeling the shadow of a BH, where it is illuminated by a distant screen emitting into a hemisphere, \cite{1}. As it was shown, in this case:
\begin{itemize}
\item The BH behaves like a \textit{gravitational lens} for rays directly coming from the screen to the observer, leading to the formation of a bright halo ring outside the dark shadow of the BH.
\item Inside the shadow, fainter {\it photon rings} (see \cite{mikejohnson}) appear, formed by photons that have circled the BH several times. The brightness of the rings depends on the spin of the BH and the inclination of the BH's rotation axis to the line of sight.
\end{itemize}

        In this paper, which follows \cite{1}, the photon-emitting screen is a simplified model of an accretion disk with some admissions that facilitate the construction of the shadow. The plane of symmetry of the screen coincide the equatorial plane of the BH. The screen is massless, non-rotating, and optically thick. The envelope of the vertical cross-section of the screen can be described by a parabola. The screen temperature depends on the radial coordinate $r$, which leads to new properties of BH image. As in paper~\cite{1}, the system of six ordinary differential equations was solved to construct photon trajectories (see also~\cite{40} and references therein). Note that the method used forderiving and numarically integrating the equations of null geodesics was proposed earlier in paper~\cite{41,a1}.
        
        In formulation of the problem this work is similar to the pioneer paper \cite{a2}, where the problem of the shadow shape of a Schwarzschildian black hole illuminated by a thin disk was also considered.Overthe past yearsin construction of BH shadows -- not only with the Schwarzschild metric, but also in more general cases -- new results have been obtained (see the review \cite{a3} and references therein). Modern computational capabilities allow to trace complicated trajectories of photons (see \cite{a4}), and thus to model detailed shadow images and to construct maps od relative intensities of different parts of the images for the subsequent use of the maps in the determination of black hole masses and spins.

\section{Parabolic screen}

        Assuming a massless screen, the space-time metric around the BH has the form:
\begin{equation}
ds^2=K\text{d}t^2-\frac{\text{d}r^2}{K}-r^2\left(\text{d}\theta^2+\sin^2\theta \,\text{d}\varphi^2\right),\quad K\equiv 1-\frac{r_{S}}{r},
  \label{BH_metric}
\end{equation}
where $r$ is the radius of curvature of the 2-sphere with coordinates $\theta\in(0,\pi)$ and $\varphi$, $r_{S}$ is the Schwarzschild radius, $\theta=\pi/2$ is screen symmetry plane.

\begin{figure}[t]
\centering
\includegraphics[width=0.8\textwidth]{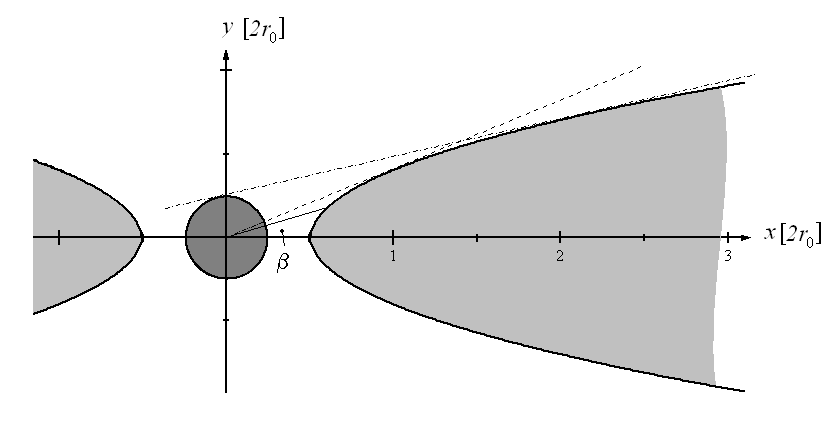}
\caption{Diagram of an emitting screen in the vicinity of a black hole (BH). The BH is located at the origin of coordinates, the dark gray zone corresponds to the region of space bounded by photon orbits in the minimum orbit for the case of a Schwarzschild BH.The light gray area corresponds to the inner area of the screen. In this picture $r_{ls}= r_0/2$, $R_{ls}= 1/4$, $b=1/2$.}.
\label{fig1}
\end{figure}

        Let us consider the surface of an axisymmetric torus-like cloud, which has the shape of a parabola in a vertical cross-section (see~Fig.~\ref{fig1}). This shape of the emissive surface allows to naturally cut off the photons generated at large radii. There is a relationship between the coordinates $x$ and $y$, describing the position of a point on the parabola:
\begin{equation}
x=\frac 12\left(1+\frac{y^2}{b}\right),
\label{xy}
\end{equation}
where $b$ is the parameter of the parabola, related to the value of its "opening". The coordinates $x$ and $y$ are measured in units of $2r_0$ and can be written as $x=R\sin\theta$ and $y=R\cos\theta$, where $R=r/2r_0$ is the dimensionless radius, $r_0$ -- radial coordinate of the inner edge of the screen.
\begin{equation}
r_0\geq 2r_{ls},
\label{slo}
\end{equation}
where $r_{ls}=3r_S/2$ is the minimum radius of the circular orbit for photons, along which they can revolve around the BH. Photons that have passed inside the light surface fall on the BH and cannot turn towards the observer (dark circle in Fig.~\ref{fig1}). The area between the BH and the screen is completely transparent\footnote{This assumption is of a model nature and may not be fulfilled depending on the physical conditions near the BH (structure of accretion flows), see the discussion in \cite{a3}. }.

        Since the screen coordinates consist only of the BH 2-sphere metric variables (see eq.~(\ref{BH_metric})), the shape of the screen (this is the envelope of the zone shaded in light gray in Fig.~\ref{fig1}) can be parameterized as follows :
\begin{equation}
\begin{array}{ll}
 R\cos\beta & =B-b, \\
 R\sin\beta & =\pm\sqrt{2b\left(B-\frac 12-b\right)},
\end{array}
\label{Rcos}
\end{equation}
where $\beta=\pi/2-\theta\in(-\pi/2,\pi/2)$, $B=\sqrt{R^2+b+b^2}$.
In the curved space-time of a BH, these equations relate the angular coordinate of any point of the screen profile above the central plane of the disk to the radius of curvature at this point, the radial coordinate $r$, and the disk thickness parameter $b$, i.e. $\beta=\beta(R,b)$. In this case, the coordinates of the Cartesian reference system~$(x,y)$ associated with the BH can be used as the coordinates of the radius-vector $\vec R=(x,y)$ with $x=R\cos\beta$ and $ y=R\sin\beta$, for which the relations $R=\sqrt{x^2+y^2}$, $B=x+b$, etc.

\section{Effect of the parabolic screen on the structure of the BH shadow}

       The designed model specifies which parts of the screen are responsible for the brightness of different parts of the BH image depending on the value of the thickness parameter $b$ and the temperature distribution of the screen $T(R)$.

       The maximum screen angle above the $y=0$ plane (see Fig.~\ref{fig1}) grows with the thickness parameter, $\vert\beta\vert\leq\beta_0\equiv\arctan\sqrt b$. Its value is a solution to the equation
\begin{equation}
\frac{\partial\beta}{\partial R}=0\,,
\end{equation}
which has the form 
\begin{equation}
x=1,\quad \vert y\vert=\sqrt{b}=\tan\beta_0,\quad R=\sqrt{1+b}=\frac{1}{\cos\beta_0}\,. 
\end{equation}
Hereafter, the part of the screen with $\vert x\vert\leq1$ will be called the edge of the screen, and the rest of it with $\vert x\vert>1$ will be called the actual screen. The screen will be considered extended along the radial coordinate.

         One should distinguish between the actual image of a BH, formed by photons subjected to strong gravitational lensing of the BH field, and the image of a disk, formed by photons that came to the observer ``directly'', i.e. those for which the effect of gravitational lensing is weak. The image of the latter will be called ``halo''. Thus, the boundary between the shadow and the halo (external with respect to the shadow) is formed by photons of direct radiation from the edge of the disk, slightly lensed by the BH due to small deviations in the trajectories of rays traveling from the open part of the screen to the observer. In this sense, the presence of a dark region in the central part of the screen, which can also be called a ``shadow'' by external signs, is not associated with a BH, because it's a property of the screen itself\footnote{There might not be a BH inside the shadow. The gravitational field of a BH cannot significantly change the brightness of the halo and the boundary of the shadow: halo rays come from the open part of the screen to the observer regardless of the BH, all other photons passing through a strong gravitational field, turning around the BH, cannot reach the observer as halo rays from -- for the shape of a parabolic screen.}. The presence of a BH can only be judged by the inner rings of the image formed by photons that bypass the BH in the light sphere one or more times with a corresponding decrease in the radius and brightness of the rings.

        The image of BH embedded in ``thick'' disk with $\sqrt{b}\geq l \simeq 2R_{ls}=r_{ls}/r_0$ is formed by photons from the edge of the screen ($x\leq 1$ ). All photons of the screen with $x\leq 2l^2/b$ participate in the formation of images of BHs with ``thin'' disks with $\sqrt{b}<l\leq 1/2$. Additional photons from the disk enhance the brightness of the inner rings. These photons first move along the radius ($\varphi\simeq \text{const}$ for $x>1$) and then turn towards the observer along the BH's light sphere (where $\varphi$ can change for $x<1$ ). The maximum radius of the image photon collection area from the screen surface is inversely proportional to the thickness of the thin disk\footnote{Let $\alpha=\beta-\arcsin(l/R)$ be the angle between the central plane of the disk and the momentum of the photon at the point of its emission. This photon comes to the observer through the gap $r\in(r_{ls},r_0)$. The maximum radius of the screen whose photons form the brighter outer circle of the image's main photonic ring is found as the roots of the equation:
$$
\frac{\partial\alpha}{\partial R}=0:\qquad \frac{x-1}{\sqrt{2x-1}}=\frac{aB}{\sqrt{B^2-c}},\quad a=\frac{l}{\sqrt b},\quad c=b+b^2+l^2,
$$
$$
\left(x-x_1\right)\left(x-x_2\right)=c\left(\frac{x-1}{x+b}\right)^2,\quad x_{1,2}=\sqrt{1+a^2}\left(\sqrt{1+a^2}\pm a\right)=\frac{\sqrt{b+l^2}\left(\sqrt{b+l^2}\pm l\right)}{b}.
$$
For $a<1$: $x\simeq x_1\simeq\frac{1}{x_2}\simeq\frac{R}{\sqrt{1+b}}\simeq\frac{\vert y\vert}{ \sqrt b}\simeq 1+a$.
For $a>1$: $x\simeq x_1\simeq R\simeq 2a^2$, $x_2\simeq\frac 12$, $\vert y\vert\simeq 2l$. The second solution ($x\simeq x_2$) refers to screen edge photons forming a fainter inner circle of the image's main photon ring.
}:
\[
x=R=\frac{2l^2}{b}\gg 1,\quad \vert y\vert=2l\leq 1\,. 
\]

         All inner rings of the image are focused near the light sphere of the BH, their total area is less than that of the halo. With a decrease in the thickness of the screen, the luminosity of the inner rings increases relative to the outer halo: photons of the rings come from the entire disk and are lensed in a strong inhomogeneous gravitational field, while only direct photons of the open edge of the disk pass into the inner part of the halo.

       Especially, it should be said about the main (first) photon ring inside the shadow: its brightness can exceed all other rings of the BH, including the halo. The rays of the main inner ring come to the observer from the back side of the disk, while turning through the angle $\sim\pi/2$. In the flat disk limit ($b\rightarrow 0$), the photons that form the inner rings of the shadow are collected from the entire screen. In this case, the angle of rotation of the rays for the first ring is limited, which leads to an increase in its brightness relative to other rings of the image. This brightening effect can be used to test the geometric models of disk accretion in a strong gravitational field. In the first step, we will test it in a numerical model of BH imaging in the thin disk limit. Next, we consider the images of a Kerr BH with the rotation axis perpendicular to the disk plane for different values of the angle between the BH spin and the line of sight.

\section{Thin screen approximation}

\begin{figure}[t]
\centering
\includegraphics[width=0.75\textwidth]{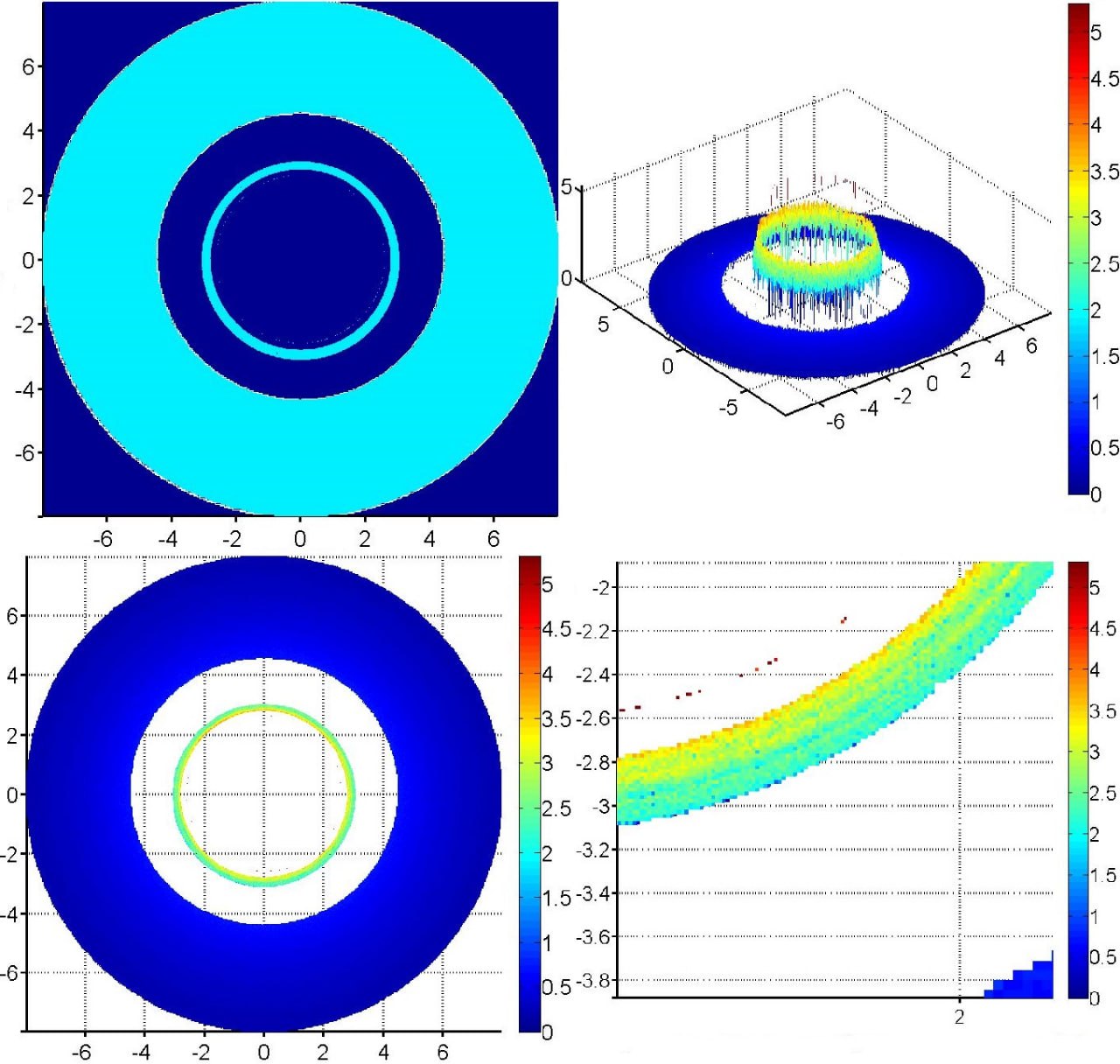}
\caption{The shadow of a Schwarzschild black hole. The angle between the screen symmetry axis and the direction towards the observer is $5^\circ$.}
\label{fig2}
\end{figure}

\begin{figure}[t]
\centering
\includegraphics[width=0.75\textwidth]{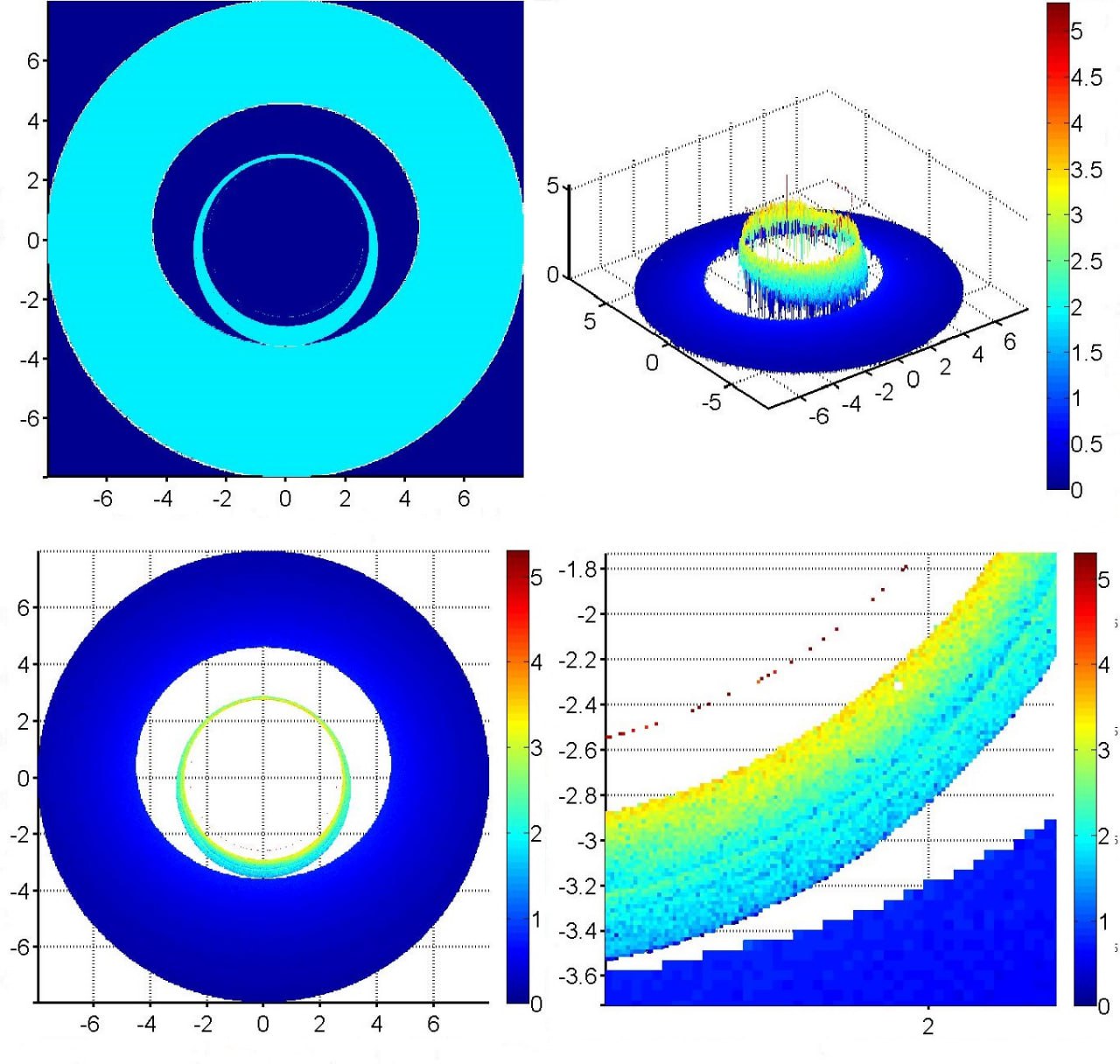}
\caption{The shadow of a Schwarzschild black hole. The angle between the screen symmetry axis and the direction towards the observer is $30^\circ$.}
\label{fig3}
\end{figure}

\begin{figure}[t]
\centering
\includegraphics[width=0.75\textwidth]{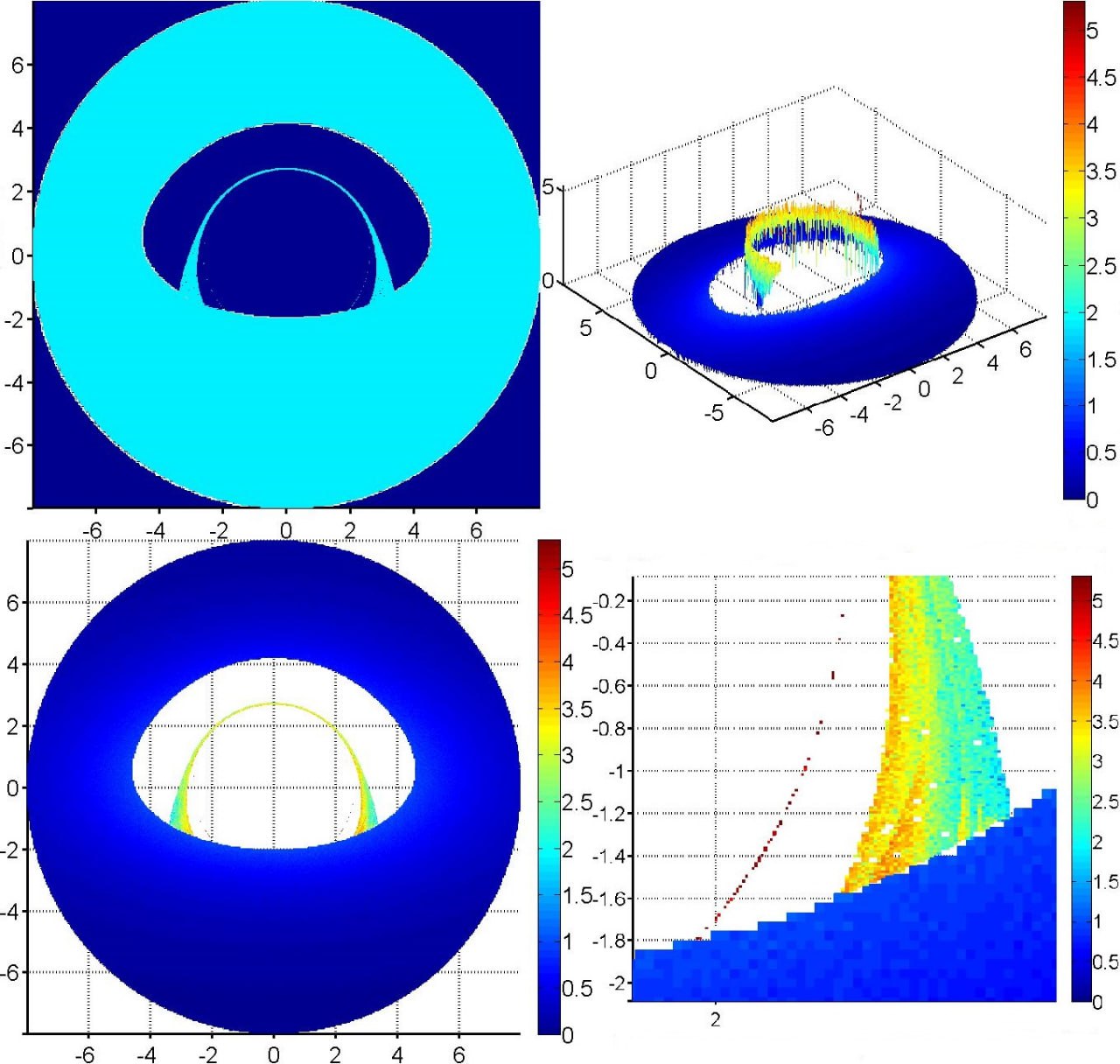}
\caption{The shadow of a Schwarzschild black hole. The angle between the screen symmetry axis and the direction towards the observer is $60^\circ$.}
\label{fig4}
\end{figure}

\begin{figure}[t]
\centering
\includegraphics[width=0.75\textwidth]{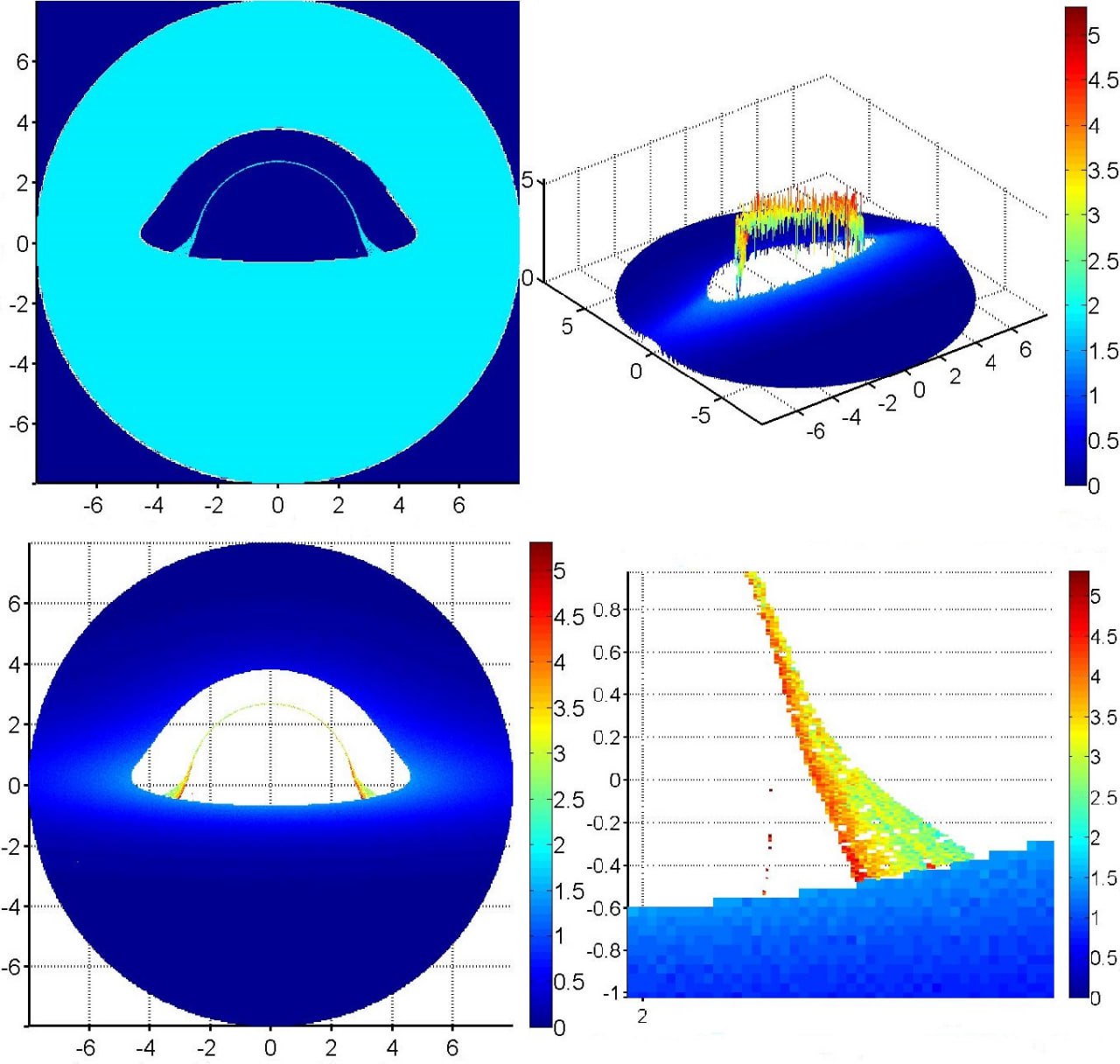}
\caption{The shadow of a Schwarzschild black hole. The angle between the screen symmetry axis and the direction towards the observer is $80^\circ$.}
\label{fig5}
\end{figure}

In the $b\rightarrow 0$ approximation, the parabolic screen degenerates into a flat screen. As before in \cite{1}, when solving the problem of constructing geodesics, it is assumed that each point on the screen surface emits photons into the hemisphere. Following the previously developed methodology, the shadow images shown in Fig.~\ref{fig2}-\ref{fig5} were constructed.

        For each image of a BH, 4 million photon trajectories were built. It was assumed in the calculation that the screen does not rotate and that the photons do not experience a gravitational redshift. Also, it was assumed that the photons are emitted by the screen, which has a temperature whose dependence on the radius coincides with the analogous dependence for the relativistic Novikov-Thorne disk (see ~\cite{NovikovThorne}). The emissivity of both sides of the screen is identical. Some color inhomogeneities in the figures are due to numerical effects and features of standard graphics programs used to build images.

Each of the figures~\ref{fig2}-\ref{fig5} consists of four panels. A schematic representation of a shadow is given at the top left: if at least one photon comes from a pixel, the pixel is colored cian, if there are no photons, it is dark blue. The bottom left panel shows the relative intensity of radiation from different parts of the BH shadow on a logarithmic scale. On the upper right, the image from the lower left panel is artificially tilted for better visualization, the relative radiation intensity is plotted along the vertical axis on a logarithmic scale. The lower right panel shows a several times enlarged image of the area with photonic rings. Fig.~\ref{fig2}-\ref{fig5} differ in the angle from which the observer sees the screen. In the first case (Fig.~\ref{fig2}), the angle between the screen's symmetry axis and the direction towards the observer is five degrees, in the next three figures it is 30, 60 and 80 degrees.

        For theoretical analysis, the upper left panel of figures is of particular interest, and for studying the possibilities of observing constructed images, the lower left panel is of particular interest.

        On the upper left panel of fig.~\ref{fig2} one can visually distinguish three rings of different widths. The outer wide ring is formed by the photons coming from the screen ``directly'' i.e. emitted by the side of the screen facing the observer. In aforementioned terms, we call it ``halo''. The inner edge of the halo has the coordinate $x=1/2$ (see Fig.~\ref{fig1}). The middle ring is formed by the photons subjected to a strong gravitational lensing and passing through the gap between the light surface of the BH and the inner edge of the screen, i.e. these are photons emitted from the back of the screen, which would not be visible were it not for gravitational lensing.This is the first (main) photon ring. Finally, the inner ring is formed by the photons emitted by the side facing the observer, but having circled around the BH, i.e. the angle of their rotation in the field of the BH $\sim 3\pi/2$. With an increase in the number of geodesics, as well as the scale of the figure, other rings become noticeable, formed by small number of photons that have circled the BH with even larger rotation angles.

        In order to make the image of the BH shadow closer to the real picture, it was taken into account that each pixel of the image contains a different number of photons. The lower left panel of the figures \ref{fig2}--\ref{fig5} displays the relative intensity of emission. It can be seen that for a halo it slowly decreases with increasing radius, which corresponds to a decrease in temperature for a relativistic disk. The upper right panel illustrates the difference between the intensities of the photon ring and the halo. Here it is clearly seen that their ratio is about three orders of magnitude. The lower right panel allows us to examine the photon ring in more detail and see that its outer (closer to the halo) edge has a much greater intensity than the inner one.

        As the angle between the screen symmetry axis and the line of sight increases, the photon ring approaches the inner edge of the halo, but does not merge due to the difference in intensities; the ring transforms into a sickle. At large values of the angle, the main radiation flux is concentrated in two spaced regions.

\subsection{Kerr black hole}

\begin{figure}[t]
\centering
\includegraphics[width=0.75\textwidth]{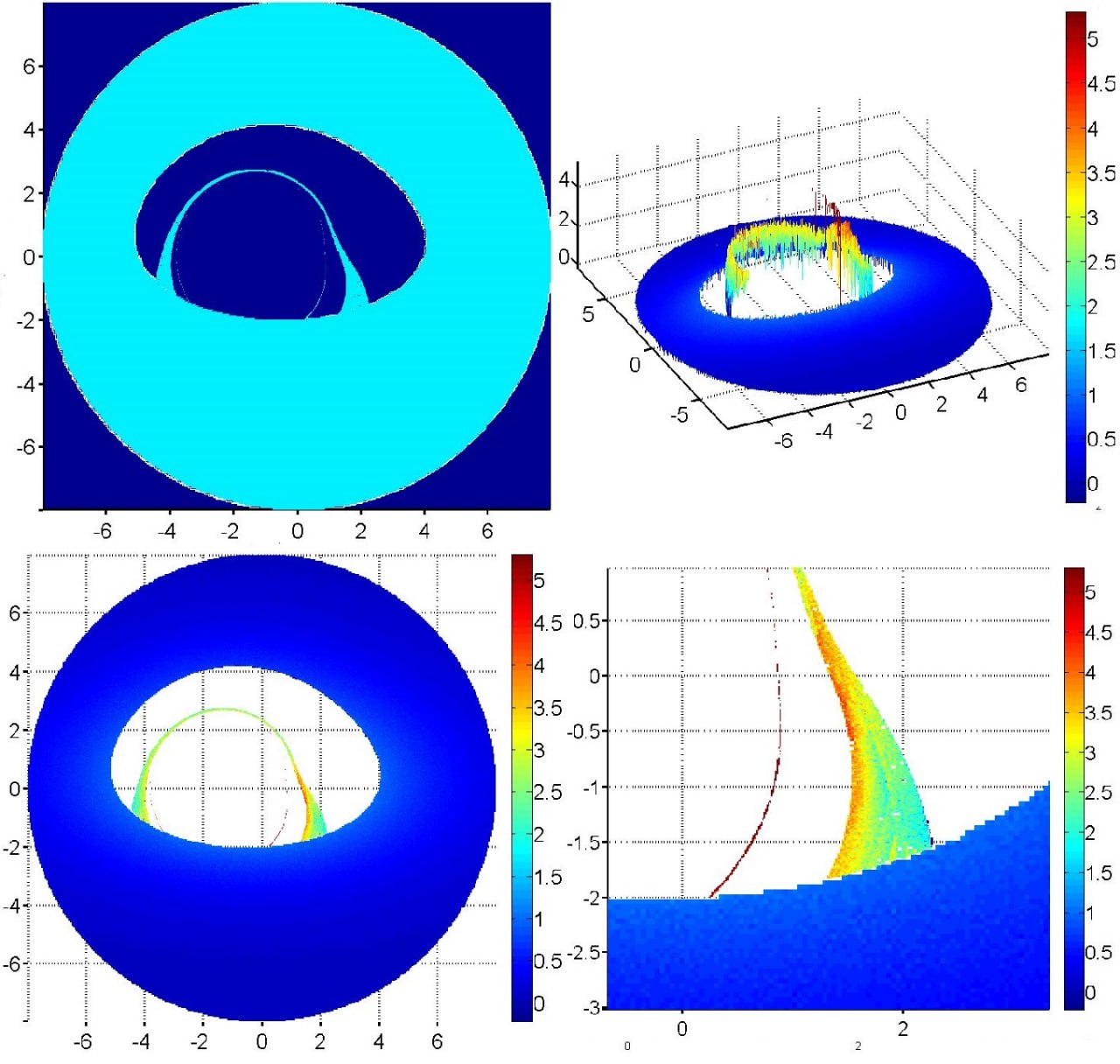}
\caption{The shadow of a Kerr black hole. The angle between the screen symmetry axis and the direction towards the observer is $60^\circ$.}
\label{fig6}
\end{figure}

\begin{figure}[t]
\centering
\includegraphics[width=0.75\textwidth]{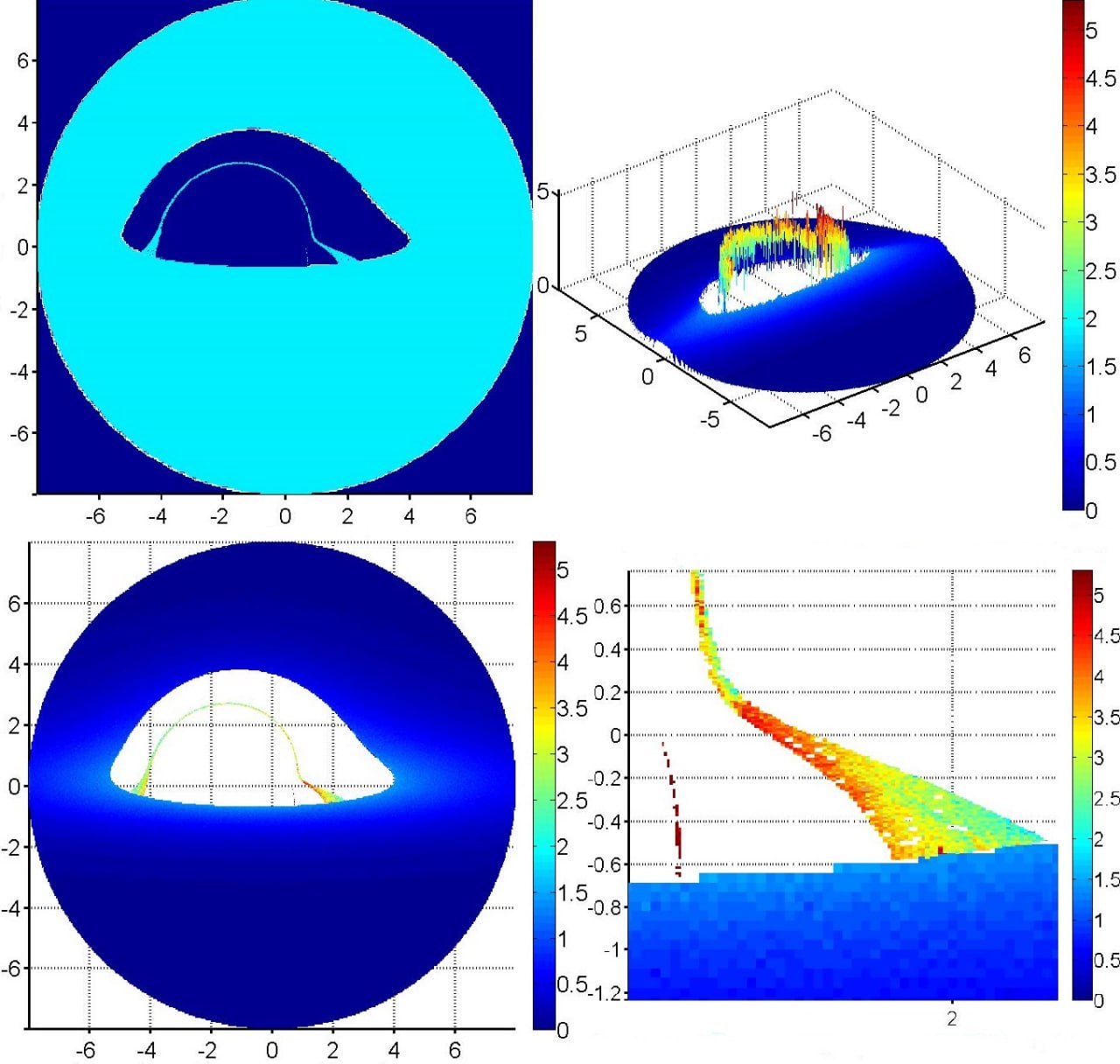}
\caption{The shadow of a Kerr black hole. The angle between the screen symmetry axis and the direction towards the observer is $80^\circ$.}
\label{fig7}
\end{figure}

        Similar calculations were carried out for a Kerr BH whose rotation axis coincides with the screen symmetry axis and whose dimensionless rotation parameter is $a=0.9982$. In the case when the angle between the axis of rotation of the BH and the direction towards the observer is small, the effects brought by the rotation are hardly noticeable. The figures~\ref{fig6} and \ref{fig7} illustrate the cases where this angle is 60 and 80 degrees, respectively. As expected, the photon ring becomes asymmetric, depending on the spin.

\section{Discussion and Conclusions}

        In recent years, many papers have been published on various aspects of the shadows of BHs. They concern the construction of photonic geodesics, methods for restoring images from interferometric observational data, which orbits are preferable for space-ground telescopes, and so on. Sometimes the issue of determining what, in fact, should be called the shadow of a BH and how well the term is used is also raised. The most general definition is a dark area in the central part of the source, where the central body would have to radiate if it had the usual properties, which is not entirely satisfactory, since it leaves ``behind the scenes'' the sources with a non-trivial spectral distribution.Without going into details, one can imagine an object that radiates mainly in the high frequency region, but is observed in the low frequency region. In this case, one can expect the appearance of a dark spot on a bright background formed by low-frequency radiation. It also seems logically not quite acceptable to call ``an image'' the absence of photons, which actually bring information about the object. Therefore, sometimes the term ``silhouette'' appears in the literature in relation to the shadow of a BH. This term was first usedin relation to the shadows of BHs in the paper \cite{a5}, but was not widely used. However, it was mentioned in the frequently cited paper \cite{a6}. The authors of the review \cite{a3} use this term to describe the structure of the image of a BH depending on the sources of photons that form the image.

        In this paper, we have considered the academic problem of forming an image of a BH surrounded by a screen (model disk). In comparision with the pioneering paper \cite{a2}, we constructed not only the general structure of the shadow (upper left panel in figures~2--7), bu also the distribution of relative intensity across the image (bottom left and upper right panels in figures 2--7), as well  as more detailed image of the photon ring (bottom right panel in figures 2--7). In contrast to what was stated in \cite{a7,a8}, our attention was focused on calculating the relative intensities of different parts of BH images, since thay are used in modeling shadows, taking into account the observation frequency, signal accumulation time and the $(u-v)$-plane coverage realized during the interferometric measurements.
 It has been shown that the boundary of the bright and dark zones is formed by photons coming from the inner edge of the model disk, and, in this sense, the boundary of the BH shadow and its size are determined by the inner radius of the screen, and not by the BH. A characteristic feature of the image, indicating the presence of a BH, are photon rings (ring), i.e. very narrow ring-shaped structures with high (relative to the screen) intensity. 
 This is consistent with the division of the BH image into the ``true'' image and the shadow as introduced in \cite{a4}.
 The resulting picture weakly depends on the temperature distribution over the screen, but significantly on the angle at which the model disk is observed.  If this angle is small, so that the observer sees the screen approximately from the pole, then the image depends weakly on the moment of rotation of the BH. If the angle is large, then the photon ring turns into an asymmetric photon crescent.

\section*{Acknowledgment}
        S.V.~Repin expresses his gratitude to R.E.~Beresneva, O.N.~Sumenkova, and O.A.~Kosareva for the possibility of fruitful work on this problem.

\end{document}